\documentclass[prX,twocolumn,superscriptaddress,amsmath,amssymb,showkeys,a4paper]{revtex4}
\usepackage[utf8]{inputenc}
\usepackage{amsmath}
\usepackage{amssymb}
\usepackage{setspace}
\usepackage{graphicx}
\usepackage{dcolumn}
\usepackage{bm}

\begin{document}

\title{Coupled opto-electronic simulation of organic bulk-heterojunction solar cells: parameter extraction and sensitivity analysis}
 
 \author{R. Häusermann}
  \email{hroger@phys.ethz.ch}
 \altaffiliation[Current address: ]{ Laboratory for Solid State Physics, ETH Zurich, 8093 Zurich, Switzerland}
\author{E. Knapp}
\author{M. Moos}
\author{N. A. Reinke}
\affiliation{Zurich University of Applied Sciences, Institute of Computational Physics, Wildbachstrasse 21, 8401 Winterthur, Switzerland}

\author{T. Flatz}
 \affiliation{Fluxim AG, Dorfstrasse 7, 8835 Feusisberg, Switzerland}
 \author{B. Ruhstaller}
 \email{beat.ruhstaller@zhaw.ch}
 \affiliation{Zurich University of Applied Sciences, Institute of Computational Physics, Wildbachstrasse 21, 8401 Winterthur, Switzerland}
 \affiliation{Fluxim AG, Dorfstrasse 7, 8835 Feusisberg, Switzerland}

\date{\today}

\begin{abstract}

A comprehensive opto-electronic device model for organic bulk-heterojunction solar cells  is presented. First the optical in-coupling into a multilayer stack is calculated. From the photon absorption profile a charge transfer (CT) exciton profile is derived. In this study we consider the Onsager-Braun mechanism to calculate the dissociation of the CT-excitons into free charge carriers.  These free charge carriers then migrate towards the electrodes under the influence of drift and diffusion. 
A general problem arising in computer simulations is the number of material and device parameters, which have to be determined by dedicated experiments and simulation-based parameter extraction. In this study we analyze measurements of the short-circuit current dependence on the active layer thickness and current-voltage curves in poly(3-hexylthiophene):[6,6]-phenyl-C$_{61}$-butyric acid methyl ester (P3HT:PCBM) based solar cells. We have identified a set of parameter values including dissociation parameters that describe the experimental data. The overall agreement of our model with experiment is good, however a discrepancy in the thickness dependence of the current-voltage curve questions the influence of the electric field in the dissociation process. In addition transient simulations are analyzed which show that a measurement of the  turn-off photocurrent can be useful for estimating charge carrier mobilities.

\end{abstract}
\pacs{85}
\maketitle


\section{Introduction}

After  the huge progress organic light-emitting diodes  have made in the last years, organic photovoltaic (OPV) devices attract more and more interest. State of the art OPV devices yield an energy conversion efficiency of around 6\% to 7\% for single junction cells \cite{Green:2009p1245} as well as tandem cells \cite{Kim:2007p882}. This is much less compared to other already established photovoltaic power conversion techniques which have an efficiency above 10\% or even above 20\% for crystalline silicon cells. Nevertheless OPV devices have several advantages like the possibility of low-cost production \cite{Kalowekamo:2009p1224}, room temperature processing and thin film structures. The latter two make it possible to use flexible substrates and thus the production of  flexible solar cells. This leads to low-cost, flexible and transportable  energy generators. Our research focuses on the influence of optical and electrical parameters on device performance.\\
The field of OPV 's can be separated into the planar-heterojunction devices where the donor and acceptor materials are deposited one after the other, mostly by evaporation of small molecules and the bulk-heterojunction (BHJ) devices, where the two organic materials are diluted in the same solvent and spin coated as one layer. The advantage of the BHJ structure is that most of the generated excitons reach a nearby donor-acceptor interface where they dissociate into free charge carriers. This efficient exciton harvesting leads to  higher power conversion efficiencies for BHJ devices.\\ 
Numerical models for organic BHJ devices provide insight into operating mechanisms and allow for device structure optimization. The simulation of OPV devices can be separated into two parts, firstly there is the in-coupling of light into a multilayer structure and secondly the extraction of charges which needs an electrical model.  The absorption of light within the multilayer structure is a crucial process and thus one of the main areas of numerical simulations. Harrison et al. \cite{Harrison:1997p1246} used analytical models and compared them to measured photocurrent action spectra. Later a transfer matrix formalism has been used by Petterson et al. \cite{Pettersson:1999p952} to calculate the absorbed optical energy within the multilayer structure. This formalism is widely used to optimize layer structures for single junction \cite{Moniestier:2008p879} as well as tandem cells \cite{Dennler:2007p988}. Though optical models are very reliable they are only able to give an upper limit of the achievable photocurrent. Therefore electrical models have been developed to study charge generation and transport losses in planar heterojunction \cite{Barker:2003p967} as well as BHJ devices \cite{Koster:2005p620,Monestier:2007p621,Deibel:2008p1009,Hwang:2008p1178}. These models differ in their choice of ingredients, definition of the boundary conditions as well as in their methods to solve the drift-diffusion equations. Coupled opto-electronic models have been introduced by Kirchartz et al. \cite{Kirchartz:2008p1208}, Lacic et al. \cite{Lacic:2005p878} and Kotlarski et al. \cite{Kotlarski:2008p983}, where they connected a drift-diffusion solver with an optical thin-film simulator. All the electrical models mentioned are 1D-models and they treat every layer as an effective medium, i.e. a homogenous layer. \\
In this study we present a comprehensive numerical model which considers optical as well as electrical effects. This device model has originally been developed for the analysis of organic-light emitting diodes \cite{ Leger:2003p1242,Ruhstaller:2001p158,Ruhstaller:2003p487} and was commercialised afterwards. This software, SETFOS \cite{SETFOS}, has been extended for the simulation of organic BHJ solar cells.


\section{Description of the numerical device model}

\subsection{Optical modeling}

\begin{figure}
\begin{center}
\resizebox{8.5cm}{!}{\includegraphics{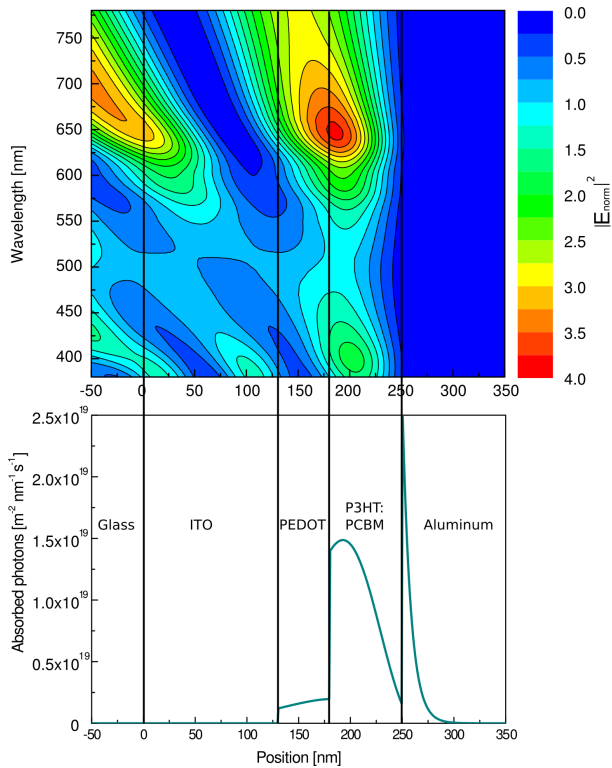}}
\end{center}
\caption{ \label{fig:field_spectrum_photon_abs} Electromagnetic field penetration plot (top) is calculated using a transfer matrix formalism. This field penetration is then used  to derive the  photon absorption rate profile (bottom).}
\end{figure}

Organic photovoltaic cells consist of a multilayer structure, with layer thicknesses on the order of the wavelength of the incident light, which is shorter than the coherence length of sunlight. Therefore an OPV is an optically coherent device. This gives rise to interference effects inside the device which can be exploited to increase the maximum photocurrent by carefully tuning the thicknesses of the individual layers \cite{Gilot:2007p609} \cite{Pettersson:1999p952} \cite{Monestier:2007p621}. An approach to find the optimal layer thicknesses from an optical point of view is to calculate the electromagnetic field inside the multilayer stack using a transfer matrix formalism. This approach has been used in the optical simulation of OLED's \cite{Leger:2003p1242}  \cite{Ruhstaller:2003p487} and OPV's \cite{Pettersson:1999p952} and is well established in the field of organic electronics. With this formalism the normalized optical electric field $E_{norm}$ is calculated throughout the multilayer stack (cf. Fig. \ref{fig:field_spectrum_photon_abs}). For plane waves, the optical field intensity within the stack is given by

\begin{equation}
I=\frac{1}{2} \varepsilon_{0}  c n |\mathit{E}|^2.
\end{equation}

\begin{figure}
\begin{center}
\resizebox{8.5cm}{!}{\includegraphics{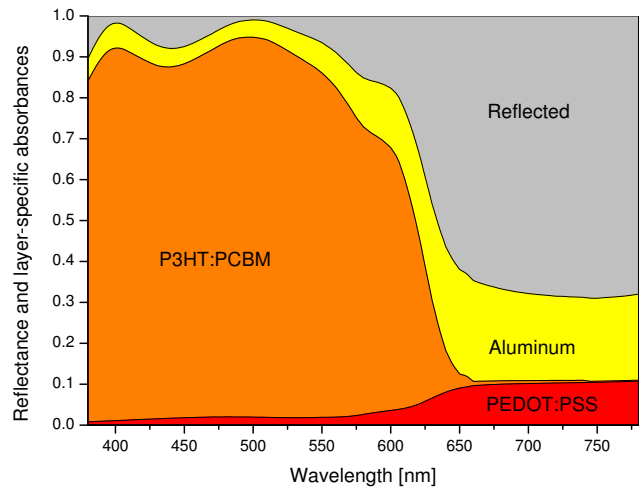}}
\end{center}
\caption{ \label{fig:layer_absorbance} Reflectance and fractional absorbance for each layer in the solar cell.}
\end{figure}

In the transfer matrix formalism we normalize the incident electric field such that the incoming electric field amplitude is 1. Therefore the electric field calculation gives the field strength relative to the field strength in the surrounding medium E$_0$. This is used to calculate the absolute energy flux by looking at the ratio 

\begin{equation}
\frac{I}{I_{0}}=\frac{\frac{1}{2} \varepsilon_{0}  c n |\mathit{E}|^2 }{\frac{1}{2} \varepsilon_{0}  c n_{0} |\mathit{E}_0|^2 } 
\end{equation}

of the illumination intensity $I_0$ in the surrounding medium and device-internally $I$. In this study the AM 1.5 spectrum is used for $I_0$. The calculation of the device-internal intensity 

\begin{equation}
 I=\frac{n}{n_{0}} |\mathit{E}_{norm}|^2 I_{0},
\end{equation}

is done wavelength and position dependent, from which the density of absorbed photons per second can be derived via

\begin{equation}
n_{photons}=\frac{\alpha I \lambda}{h c},
\end{equation}

where $\alpha$ stands for absorption coefficient which is given by $\alpha = \frac{4 \pi k}{\lambda}$ and k stands for the complex part of the refractive index (i.e. the extinction coefficient). After an integration over the illumination spectrum one ends up with a photon absorption rate profile (cf. Fig. \ref{fig:field_spectrum_photon_abs} (bottom)). The integration is done in the absorbing spectral range from 380 to 780~nm. The absorbance  of the active layer, which is the ratio of absorbed to the illumination light intensity,  drops to zero above 650~nm as is shown in Fig. \ref{fig:layer_absorbance}. It is obvious that the thickness and refractive index is required for each layer for  accurate simulation results. The n and k data were taken from Monestier et al. \cite{Monestier:2007p621}.

\subsection{Electrical modeling}

To simulate an OPV device electrically, several processes have to be taken into account. First charge-transfer-exciton generation  and dissociation, then charge drift and diffusion, and at last the charge extraction at the electrodes need to be considered. These processes will be discussed in this section.

\subsubsection{Charge-transfer-exciton dissociation}

\begin{figure}
\begin{center}
\resizebox{8.5cm}{!}{\includegraphics{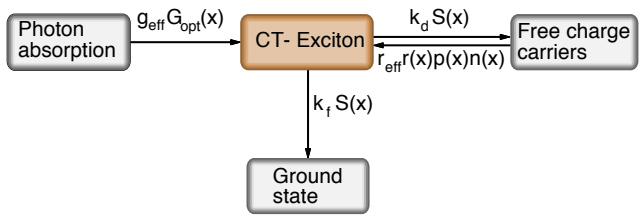}}
\end{center}
\caption{ \label{fig:CT-exciton_processes} Relevant processes for CT-exciton modeling.}
\end{figure}

The absorption of a photon generates an exciton which then diffuses to a donor-acceptor interface. At this interface the exciton dissociates and generates a charge-transfer-exciton (CT-excitons). In this model we assume that a photon directly generates a CT-exciton. Therefore only a continuity equation for CT-excitons is needed (cf. Fig. \ref{fig:CT-exciton_processes}):

\begin{equation}
\begin{aligned}
&\frac{dS(x)}{dt} =  \\
& r_{eff}r(x)p(x)n(x)-k_{f} S(x) -k_d S(x) + g_{eff}G_{opt}(x).
\end{aligned}
\label{Eqn:CT-Exc:cont}
\end{equation}

CT-exciton density $S$ is fed by two processes: Firstly an absorbed photon directly generates a CT-exciton.  $G_{opt}(x)$ stands for the photon absorption profile shown in figure \ref{fig:field_spectrum_photon_abs} (bottom), whereas  $g_{eff}$ is a photon-to-CT-exciton conversion efficiency.  We set $g_{eff}$ to zero outside of the active layer, which is P3HT:PCBM in the device under consideration here. Secondly recombination of free charge carrier pairs generates a CT-exciton. This recombination process is treated by the Langevin theory which will be discussed in the next section when the drift-diffusion model is explained. There are also two CT-exciton state depletion mechanisms $k_f$ and $k_d$ which stand for the decay of a CT-state and the dissociation respectively as described by the Onsager-Braun theory. This continuity equation for CT-excitons is coupled with the drift-diffusion model which will be discussed later.\\
CT-exciton dissociation has been studied by L. Onsager  \cite{Onsager:1934p996} \cite{Onsager:1938p994}. The initial pair separation distance is overestimated when measurements are compared with the Onsager theory. Therefore C. L. Braun \cite{Braun:1984p596} extended the theory by Onsager. In this model a finite decay rate of the CT-exciton has been considered.  The dissociation probability P is then given by

\begin{equation}
\begin{aligned}
P(T,E,k_{f},a) & = \frac{k_{d}}{k_{d}+k_{f}}\\
k_{d}(T,E,a) & =\frac{3 \mu e}{4 \pi \varepsilon a^3}e^{\left(-\frac{\Delta \mathcal{E}(a)}{k_{B} T}\right)}\frac{J_{1}\left(2 \sqrt{-2 b(T,E)}\right)}{\sqrt{-2 b(T,E)}}\\
\Delta \mathcal{E}(a) & =\frac{e^2}{4 \pi \varepsilon \epsilon_{0} a}\\ 
b(T,E) & = \frac{e^3 E}{8 \pi \epsilon \epsilon_{0} k_{B}^2 T^2}.
\end{aligned}
\label{Eqn:braun_diss}
\end{equation}

In equation \eqref{Eqn:braun_diss} the mobility is the sum of the electron and hole mobility $\mu=\mu_e + \mu_h$, $J_{1}$(x) is the Bessel function of the first kind of order 1, $a$ is the initial pair separation distance of the CT-exciton, the pair binding energy  $\Delta \mathcal{E}(a)$ is calculated under the assumption that CT-excitons have the same dependence of the binding energy on the separation distance as ion pairs. $E$ stands for the electrical field which is calculated position dependent using the drift-diffusion model explained in the next section.\\
To find an analytical model, the Onsager-Braun theory makes some simplifications. It does not consider disorder in energy levels as well as in the spatial distribution of the hopping sites which is present in an amorphous organic semiconductor. Koster et al. \cite{Koster:2005p620} and Pan et al. \cite{Pan:2000p1223} introduced spatial disorder by integrating the dissociation probability over distribution of pair separations $a$. A further simplification of the Onsager-Braun model is the assumption that the dissociation takes place in a homogeneous material, neglecting the presence of donors and acceptors. Veldman et al. \cite{Veldman:2008p1182} proposed that the mobility used in the Onsager-Braun model is higher than the effective mobility measured in the bulk. They attributed this to local crystallization of the PCBM molecules. This extension allows to model CT-exciton lifetimes in the order of 10$^{-8}$~s still having a dissociation efficiency over 90\%. This extension is not considered in our study. To analyze the effects of disorder and donor acceptor interfaces in more detail, Monte-Carlo simulations are used which are not the scope of this study  \cite{STOLZENBURG:1987p1191,ALBRECHT:1995p1185,Arkhipov:2004p1216,Offermans:2005p1184,Groves:2008p1179,Marsh:2007p1010,Yang:2008p1226,Emelianova:2008p1181}. Albeit all the simplifications, the Onsager-Braun theory is a widely used model to describe the dissociation of CT-excitons in organic semiconductors.

\subsubsection{Drift-diffusion modeling}

In the semiconductor, the following drift-diffusion equations apply:

\begin{eqnarray}
\frac{dE(x)}{dx}&=&\frac{e}{\varepsilon_{r} \varepsilon_0} [p(x) - n(x)]
\label{Eqn:drift_diff:Poi}
\\
J_e(x) &=&e \mu_e n(x) E(x) + D(\mu,T) \frac{dn(x)}{dx}
\label{Eqn:drift_diff:Cur}
\\
\frac{dn(x)}{dt}&=&\frac{1}{e}\frac{dJ_e(x)}{dx}-r_{eff} r(x)p(x)n(x)+k_d S(x).
\label{Eqn:drift_diff:Con}
\end{eqnarray}

The electrical field $E(x)$ in equation \eqref{Eqn:drift_diff:Poi}  is calculated dependent on the position $x$. Where $e$ stands for the elementary charge, $n(x)$, $ p(x)$ for the electron/hole density, $\varepsilon_{0}$ for the permittivity of free space  and $\varepsilon_{r}$ for the relative permittivity.

 Equation \eqref{Eqn:drift_diff:Cur} is the current equation for electrons. The mobility $\mu_{e}$ is shown to be independent of the electrical field, temperature or density, which is a simplified assumption used in this paper. The diffusion coefficient is given by the Einstein relation:  $D= \mu k_{B} T / q$. 

 Equation \eqref{Eqn:drift_diff:Con} is the continuity equation for electrons. This equation takes the creation, migration and recombination of the charge carriers into account.  $r(x)$ stands for the Langevin recombination $r(x)  = (\mu_{e} + \mu_{p})\frac{q}{\varepsilon_{r} \varepsilon_{0}}$. The requirement for this model to be applicable is that the mean free path $\lambda < r_{c}$, where $r_{c}$ stands for the coulombic capture radius $r_{c}=\frac{e^2}{4 \pi \varepsilon_{r} \varepsilon_{0} k T}$ which for an  average organic solid ($\varepsilon_{0}\approx$ 3 - 4) gives a coulombic radius of $r_{c} \approx 14 -19$nm. This requirement is satisfied since the  mean free path in this low-mobility disordered organic semiconductor is smaller than the coulombic capture radius \cite{Pope:1999p1114} and thus this model is applicable. The Langevin recombination efficiency prefactor, $r_{eff}$, is a factor between 0 and 1 which has been introduced by Ju\v{s}ka et al.  \cite{Juska:2006p1161} to take the reduced Langevin recombination into account. It will be discussed later in this paper.\\
For equations \eqref{Eqn:drift_diff:Cur} and \eqref{Eqn:drift_diff:Con} exist analogous equations for holes instead of electrons. The above set of equations is spatially discretised in SETFOS \cite{SETFOS} using the Scharfetter Gummel discretisation \cite{Scharfetter:1969p1241}  and solved iteratively in time using the Gummel scheme \cite{Gummel:1964p1164}.

\subsubsection{Built-in voltage}

There is a debate on the nature of the open-circuit voltage V$_{oc}$ in OPVs and currently there are several alternative explanations. There is evidence, that the open-circuit voltage depends on the highest occupied molecular orbit (HOMO) and lowest unoccupied molecular orbit (LUMO) difference of the donor and acceptor molecules at the interface \cite{Brabec:2001p917} \cite{Brabec:2002p953}. Recently, this mechanism has been looked at in more detail and a linear dependence of the V$_{oc}$ on the energy of the charge-transfer absorption has been found \cite{Vandewal:2008p1170}. The open-circuit voltage also depends on the work-function of the electrodes \cite{Kawano:2006p918}. This influence seems to be more pronounced in bulk heterojunction than in planar heterojunction device setups  \cite{Uhrich:2008p990}.  Furthermore the open-circuit voltage was observed to be intensity  and temperature dependent \cite{Koster:2005p965}. This complex behavior is challenging for the simulation of these devices, because there is not a single theory which is capable of describing all of those dependencies. The approach used here assumes that every device has a built-in voltage V$_{bi}$ which is derived from the difference in the work functions of the two electrodes. This is then used to calculate the effectively applied voltage V$_{eff}$. The electric field distribution inside the device is obtained through integration of the Poisson equation (cf. Eq. \eqref{Eqn:drift_diff:Poi}) using the effective bias V$_{eff}$ as a constraint to determine the integration constant,

\begin{equation}
V_{eff} = V_{appl}-V_{bi} = \int_{0}^L E\,dx,
\end{equation}

where V$_{appl}$ is the experimentally applied voltage. Therefore, the electric field due to space charges is superimposed on the applied field.

\subsubsection{Charge extraction}

When charges reach the organic-metallic interface they are extracted from the device. The current at such an interface has been described by Scott et al.   \cite{Scott:1999p597} as a balance of injection and surface recombination. This model considers the barrier reduction at an organic-metal interface due to the electric field and the image charge potential and calculates the net injection current.

\subsubsection{Validation of the simulator}

The model presented has been validated by varying the input parameters and analyzing their influence on the output. We have observed, that the open-circuit voltage increases slightly with illumination intensity. It also depends on the mobility of the charge carriers. Furthermore it is not equal to the built-in voltage which is assumed to be the difference between the work-functions of the electrodes in this model. The fill factor is influenced by recombination losses and thus the mobility of the charge carriers and the Langevin recombination efficiency $r_{eff}$. The short-circuit current depends linearly on the illumination intensity until recombination losses take over and the increase becomes sub-linear. The results of this validation correspond to experimental observations.


\section{Estimation of the dissociation rate}

After describing the coupled opto-electronic numerical device model in the last section, it will be applied in this section to simulate an OPV device and extract  unknown CT-exciton dissociation input parameters. The layer structure is shown in figure \ref{fig:field_spectrum_photon_abs}. The indium tin oxide (ITO) layer has a thickness of 130~nm, it is followed by a 50~nm thick Poly(3,4-ethylenedioxythiophene):poly(styrenesulfonate) (PEDOT:PSS) layer. The active layer is modeled as a 1:1 weight ratio P3HT:PCBM blend which is covered by 100~nm Aluminum. The parameters for the electrical simulations which are taken from literature are shown in table \ref{tab:parameters1}. The two mobilities are taken  from Mihailetchi et al. \cite{Mihailetchi:2006p883}, who have measured the constant mobilities of electrons and holes in a P3HT:PCBM bulk heterojunction solar cell depending on the annealing temperature. The measurements used here are for samples which have been annealed at 140$^\circ$C for 5 minutes and the mobilities have been chosen accordingly.

\begin{table}
\begin{center}
\begin{tabular}{lcrlr}
Description & Parameter              & Value &   Unit \\
\hline
Workfunction PEDOT	&& 5 & eV\\
Workfunction Al	&& 4.31 & eV\\
HOMO P3HT \cite{Kim:2006p613} &&	5.3& eV\\
LUMO PCBM \cite{Kim:2006p613} &&	3.7& eV\\
Electron mobility \cite{Mihailetchi:2006p883}  & $\mu_{n}$ & 5$\times$10$^{-7}$ &m$^2$/Vs    \\
Hole mobility \cite{Mihailetchi:2006p883}          & $\mu_{p}$ & 1$\times$10$^{-8}$ &m$^2$/Vs    \\
\end{tabular}
\caption{Input parameters for the thickness  dependent simulation of figure \ref{fig:thickness}.}
\label{tab:parameters1}
\end{center}
\end{table}

\begin{figure}
\begin{center}
\resizebox{8.5cm}{!}{\includegraphics{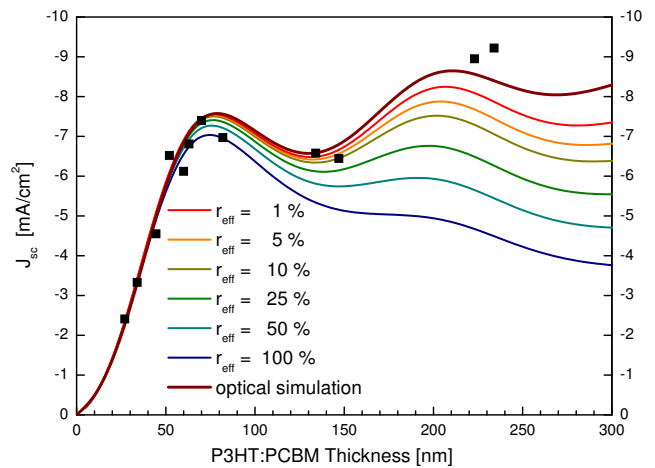}}
\end{center}
\caption{ \label{fig:thickness} Thickness dependence of the short circuit current J$_{SC}$ with varying Langevin recombination efficiency $r_{eff}$. A purely optical simulation is also shown. The simulations are compared with measurements ($\blacksquare$) by Gilot et al. \cite{Gilot:2007p609} }
\end{figure}

The unknown parameters are the decay rate $k_f$, the pair separation distance $a$ and the photon-to-CT-exciton conversion efficiency $g_{eff}$ which have been introduced in equation \eqref{Eqn:CT-Exc:cont}  and \eqref{Eqn:braun_diss} respectively. These parameters can be determined by applying the device model to thickness dependent measurements of the short-circuit current. Some reports in literature have used a simplified OPV model without CT-exciton dynamics (cf. Ju\v{s}ka et al. \cite{Juska:2006p1161}) and have found reduced recombination rates. Similarly our first analysis step assumes that absorbed photons directly generate free electron hole pairs and that recombining charges are lost and not fed into the continuity equation for CT-excitons \eqref{Eqn:CT-Exc:cont}. This reduced model thus consists of equations \eqref{Eqn:drift_diff:Poi}, \eqref{Eqn:drift_diff:Cur}  and the modified version of equation \eqref{Eqn:drift_diff:Con} which reads as follows

\begin{eqnarray}
\frac{dn(x)}{dt}&=&\frac{1}{e}\frac{dJ_e(x)}{dx}-r_{eff} r(x)p(x)n(x)+g_{eff}G_{opt}(x),
\label{Eqn:drift_diff:ConSimp}
\end{eqnarray}

where $g_{eff}$ stands for the photon-to-electron conversion efficiency. \\
Figure \ref{fig:thickness} shows a purely optical simulation (top curve) to calculate the maximum achievable photocurrent depending on the device thickness. The other six curves are for coupled opto-electronic simulations where the Langevin recombination efficiency $r_{eff}$ has been varied between 1 and 0.01. These simulation are compared with measurements by Gilot et al. \cite{Gilot:2007p609}. For devices with an active layer thickness below 50~nm electrical losses can be neglected. Therefore the photon-to-CT-exciton  conversion efficiency $g_{eff}$ is  fitted to this part of the curve.  A value of $g_{eff}$= 0.66 has been determined which is consistent with the analysis by Gilot et al. \cite{Gilot:2007p609} on the same thickness dependent data. For thicker devices electrical losses play an important role. Due to measurement uncertainties it is hard to give a final value, but figure \ref{fig:thickness} suggests that  the Langevin recombination efficiency $r_{eff}$ in the simplified model is 10\% or lower. 
In the full model there is an equilibrium between CT-exciton formation and dissociation. We translate the observation of the reduced $r_{eff}$ in the simple model to the dissociation probability in the full model, since the rate $r \times r_{eff}$ (simple model) corresponds to $r \times (1-P)$ (full model) (cf. with equations  \eqref{Eqn:braun_diss} \eqref{Eqn:drift_diff:Con} ). Thus $P$ must be 90\% or higher.\\

\begin{figure}
\begin{center}
\resizebox{8.5cm}{!}{\includegraphics[angle=90]{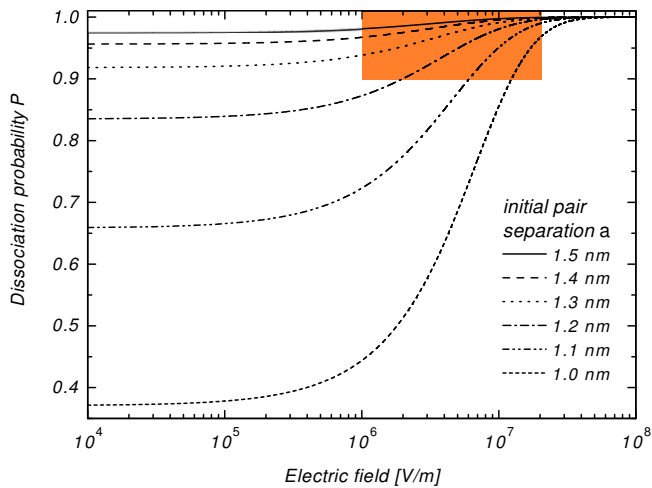}}
\end{center}
\caption{ \label{fig:braun_disseff} Dissociation probability according to the Onsager-Braun theory depending on the electrical field for several initial pair separation distances $a$. The decay rate has been kept fixed at $k_f$~=~$1\times 10^5$~s$^{-1}$. }
\end{figure}

\begin{figure}
\begin{center}
\resizebox{8.2cm}{!}{\includegraphics{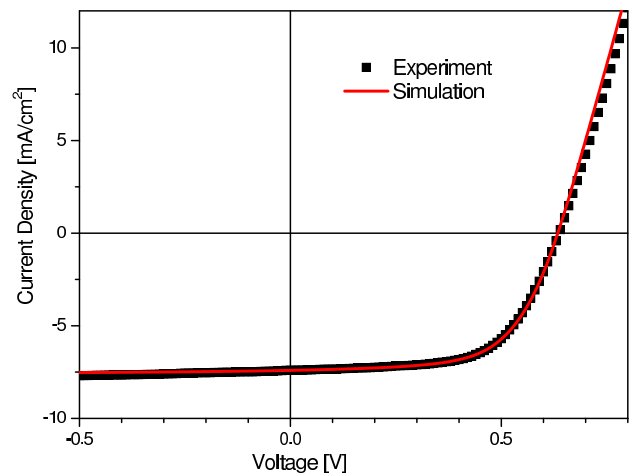}}
\end{center}
\caption{ \label{fig:JV_fit} Comparison of a simulated and experimentally measured current-voltage curve for an active layer thickness of 70 nm. The parameters used for this simulation are summed up in table 1 and 2. }
\end{figure}

With this information it is possible to determine the unknown CT-exciton parameters by analyzing the Onsager-Braun model of CT-exciton dissociation. The initial pair separation distance $a$ can be determined under the assumption that $k_f$ is set to $1\times 10^5$ s$^{-1}$. The current density in figure  \ref{fig:thickness}  has been measured and calculated under short-circuit current conditions. From the open-circuit voltage which is about 0.6 V and the device thickness the internal electric field is estimated to be in the range of 1$\times$10$^6$ to 2$\times$10$^7$ V/m for thick and thin devices respectively. Figure \ref{fig:braun_disseff} shows the desired dissociation efficiency (over 90\%) and the internal electrical field range together with different initial pair separation distances which have been varied between 1.5~nm and 1~nm. From this plot alone the only condition for $a$ is that it has to be larger than 1.2~nm. The best fit value for the pair separation distance $a$ has been chosen to be 1.285~nm, by comparing experimental current-voltage curves with simulated curves for an active layer thickness of 70~nm (cf. Fig. \ref{fig:JV_fit}). The thickness of 70~nm has been chosen because it is close to the first peak of the short-circuit current. Now, all the parameters for the full model are determined and summarized in table \ref{tab:parameters1} and \ref{tab:parameters2}. They are used in the following simulations where the full model including CT-exciton dissociation is considered. 

Note that in the full model there is no need to reduce the recombination efficiency. Even though the original recombination rate is used (i.e. $r_{eff}$ = 1), there are small recombination losses due to the competing dissociation. Also note that the value of $g_{eff}$ which was determined in the simple model is also applicable in the full model since with either model the parameter extraction is done by dividing the calculated theoretical maximum of the photocurrent by the measured one, preferably for thin devices were transport losses can be neglected.

\begin{table}
\begin{center}
\begin{tabular}{lcrlr}
Parameter & Symbol              & Value &  Unit \\
\hline
Recombination eff. & $r_{eff}$ & 1&   \\
Decay rate 	  & $k_{f}$&  1$\times$10$^{5}$ &s$^{-1}$      \\
Pair separation & $a$ & 1.285 &nm      \\
Optical generation eff. & $g_{eff}$	&0.66 &\\
\end{tabular}
\caption{Derived simulation parameters from figures \ref{fig:thickness},  \ref{fig:braun_disseff} and \ref{fig:JV_fit}}
\label{tab:parameters2}
\end{center}
\end{table}

\section{Sensitivity analysis}

In the previous section a set of parameters has been found  (cf. table \ref{tab:parameters1} and \ref{tab:parameters2} ) such that the simulation fits the measurements. In this section the sensitivity of the model on the input parameters is investigated. This sensitivity analysis gives insight into simulation as well as experiment: First we analyze and study which parameters influence the outcome of the simulation, secondly the sensitivity analysis also helps to identify experimental setups where the influence of different parameters can be separated.\\
The sensitivity analysis has been conducted as follows: Firstly, a reference has been simulated with the parameters indicated in table  \ref{tab:parameters1} and \ref{tab:parameters2}. Secondly, five input parameters have been chosen, Langevin recombination efficiency $r_{eff}$, the electron/hole mobility $\mu_n$/$\mu_p$, the pair separation distance $a$ and the decay rate $k_f$.  The first three parameters ($r_{eff}$, $\mu_n$, $\mu_p$) have been varied $\pm$ 10\% and  because the influence of the CT-exciton parameters ($a$, $k_f$) is larger, these two parameters have been varied by 3\% only. Therefore in this step ten simulations have been done (five parameters with variation in two directions each). As a last step, the reference curve has then been subtracted from the variated curves. This difference between the reference curve and the varied one has then been plotted for each parameter. This procedure reveals possible orthogonalities among the parameters.

\subsection{Thickness dependent sensitivity}

\begin{figure}
\begin{center}
\resizebox{8.3cm}{!}{\includegraphics{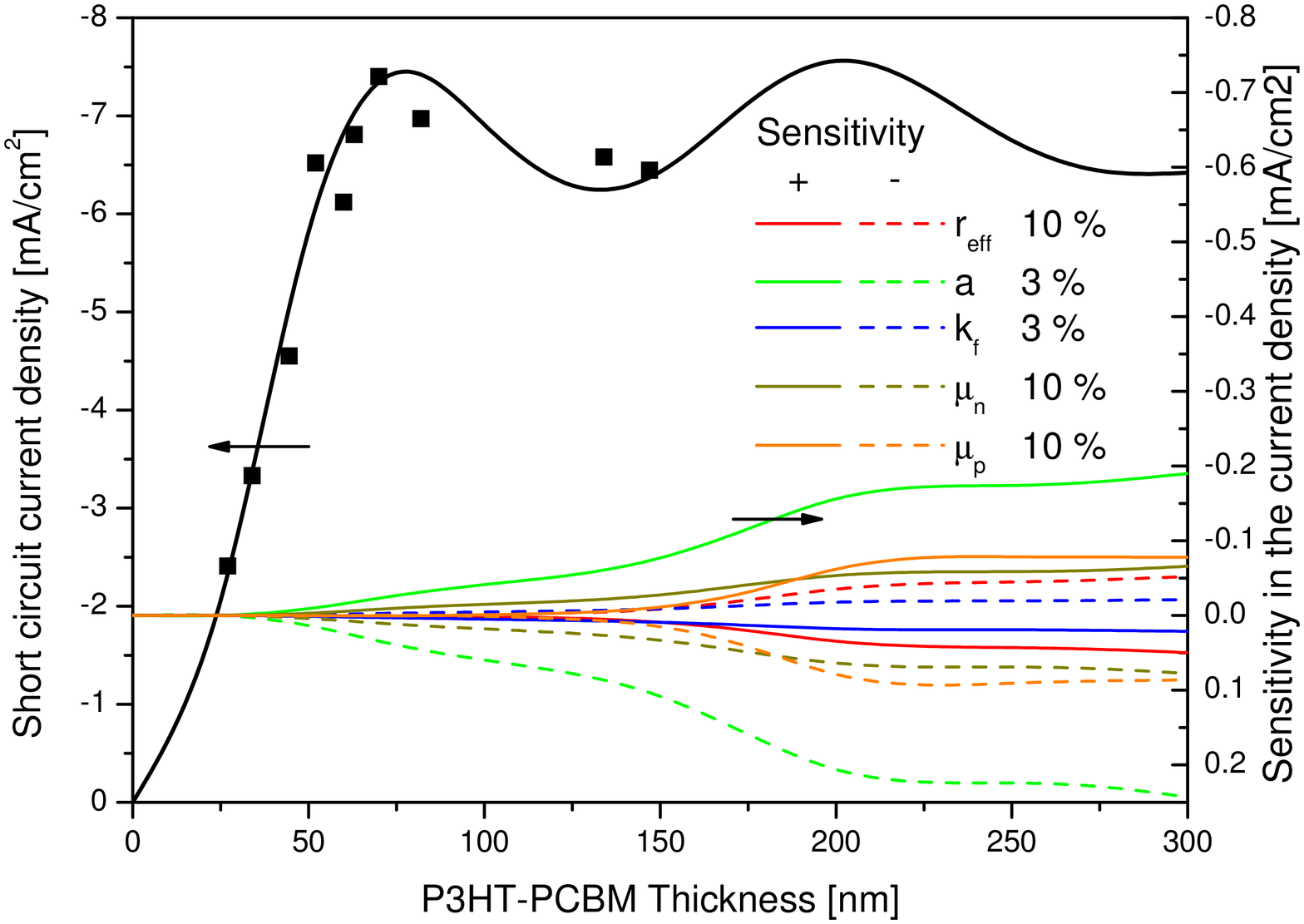}}
\end{center}
\caption{ \label{fig:thickness_Sens} Sensitivity  of the current density for a thickness dependent simulation.}
\end{figure}

First of all the thickness dependence of the short-circuit current has been analyzed, the results are shown in figure \ref{fig:thickness_Sens}. The measurements are also shown, and agree the simulation for active layer thicknesses up to 150~nm. The two measurements around 230~nm which are shown in figure \ref{fig:thickness} are lying outside the axis scale in figure \ref{fig:thickness_Sens} and can not be modeled with the parameter set used in this study. This suggests that the different spin-speed and solvent concentrations that have been used to achieve the higher active layer thicknesses \cite{Gilot:2007p609} also changed electrical properties of the active layer and therefore can not be modeled with a single set of parameters.\\

As is shown in figure  \ref{fig:thickness_Sens}, the influence of the electric input parameters are only relevant for an active layer thickness larger than about 50~nm, which has already been seen for the Langevin recombination efficiency in figure  \ref{fig:thickness}. The pair separation $a$ has the highest influence on the short-circuit current although it has been varied by only 3\%. The influence of all the parameters grows with increasing active layer thickness.  \\
Further analysis, not shown here, reveals a strong linear dependence between the three parameters  pair separation distance $a$, decay rate $k_f$ and electron mobility $\mu_n$.  These three parameters are all involved in the dissociation of CT-excitons (cf. Eq. \eqref{Eqn:braun_diss}). The thickness variation influences the electric field inside the device. Therefore the electric field dependence of the CT-exciton dissociation is probed. The three parameters which show the same thickness dependence are all pre-factors for the field dependence as shown in Eq. \eqref{Eqn:braun_diss}. In these equations the hole mobility is added to the electron mobility to determine the dissociation rate $k_d$, but in the device investigated here the electron mobility is higher by a factor of 50 than the hole mobility and therefore dominates the dissociation. The hole mobility $\mu_p$ and the Langevin recombination efficiency $r_{eff}$ both have a unique thickness dependence, which is slightly different from each other.

\subsection{Current-voltage curve sensitivity}

\begin{figure}
\begin{center}
\resizebox{8.5cm}{!}{\includegraphics{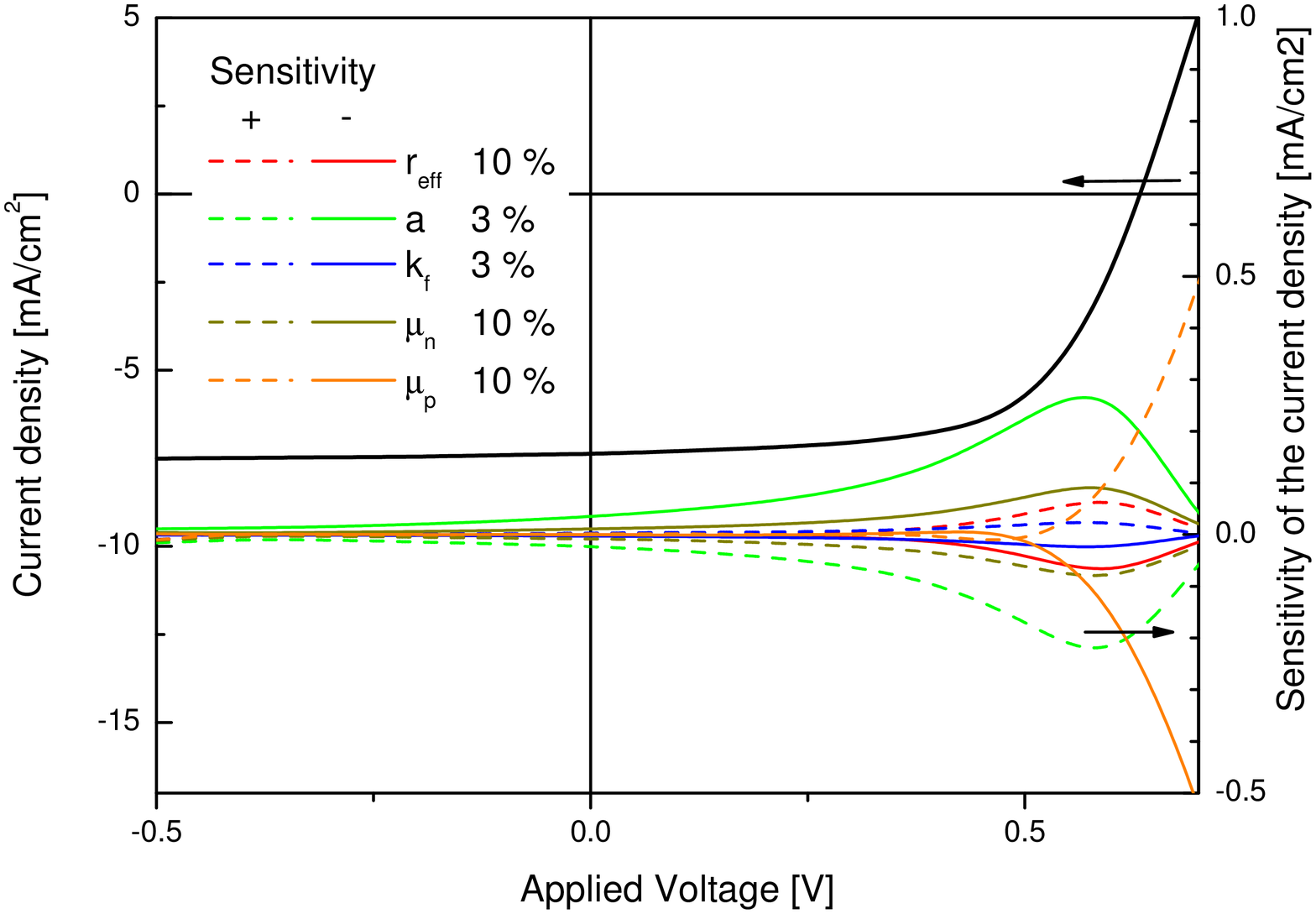}}
\end{center}
\caption{ \label{fig:IV_Sens} Sensitivity  of the current density in for a current-voltage curve.}
\end{figure}

Figure  \ref{fig:IV_Sens} shows the sensitivity analysis for a current-voltage curve with an active layer thickness of 70~nm. The current-voltage curve is the same one as in figure \ref{fig:JV_fit}. The sensitivity analysis clearly demonstrates that all parameters have the highest influence on the 
current-voltage curve in the fourth quadrant and close to the open-circuit voltage V$_{oc}$. Again the pair separation distance $a$ has the strongest influence on the output. The hole mobility clearly has a different influence on the output than all the other parameters. Especially for applied voltages which are higher than V$_{oc}$ the hole mobility has the highest influence on the current. This suggests that the hole mobility is the limiting factor for the current (photocurrent plus injected current) drawn from the device at higher voltages. 

\begin{figure}
\begin{center}
\resizebox{8.2cm}{!}{\includegraphics{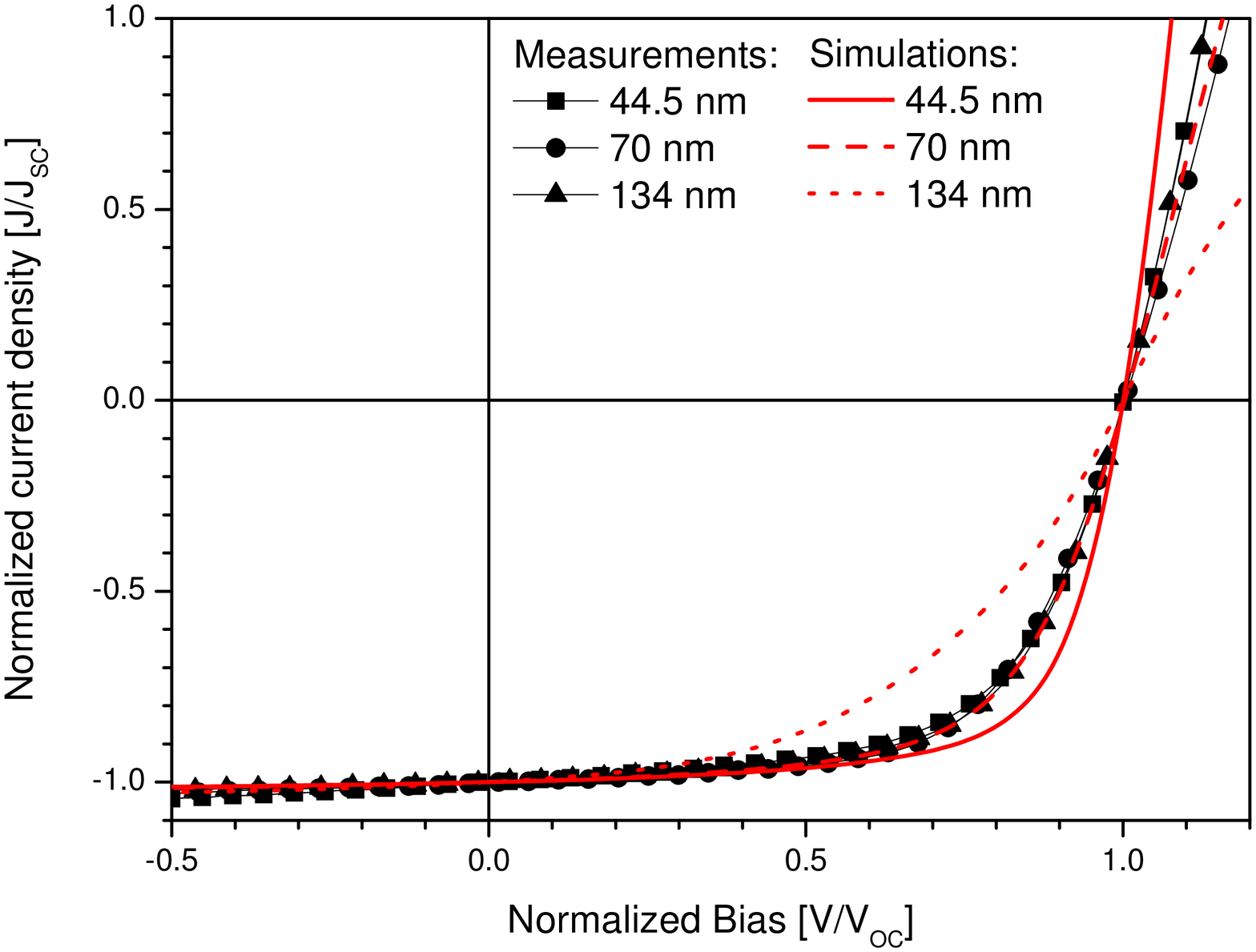}}
\end{center}
\caption{ \label{fig:IV_normalized} Normalized comparison between measured and simulated J-V curves for different active layer thicknesses.}
\end{figure}

Further analysis (not shown) reveals that for the current-voltage curve under consideration there are only two clearly linearly independent sets of parameters. The first set comprises the pair separation $a$, the decay rate $k_f$, the electron mobility $\mu_n$ and the Langevin recombination efficiency $r_{eff}$, the second one is the hole mobility $\mu_p$. Although the Langevin recombination has a slightly different voltage behavior than the other three parameters in this group it is very similar. This is interesting because in the sensitivity analysis of the thickness dependent short-circuit current shown in figure \ref{fig:thickness_Sens} the influence of the Langevin recombination efficiency was more similar to the  one of the hole mobility. \\
For all the measured short-circuit current points in  figure \ref{fig:thickness} a current-voltage curve was measured as well. An analysis of the influence of the device thickness on the current-voltage curve is performed for three representative thicknesses which is shown in figure \ref{fig:IV_normalized}. The short-circuit current as well as the open-circuit voltage depend on the amount of the absorbed photons. Therefore to compare the current-voltage curves for different thicknesses the short-circuit current as well as the open-circuit voltage need to be normalized to exclude optical effects. The comparison in figure \ref{fig:IV_normalized} shows, that the shape of the experimental current-voltage curve does not change for different thicknesses of the active layers. In contrast the simulated current-voltage curves show a clear dependence of the shape on the device thickness. This indicates that in the fourth quadrant the influence of the electric field on dissociation efficiency and thus recombination losses is overestimated in the device model presented here. A possible explanation for this discrepancy might be the nature of the Onsager -Braun dissociation mechanism, which does not consider energetic disorder. Though much used, the Onsager-Braun model is not necessarily the best physical description of charge separation and more refined models have been developed. For instance Wojcik et al. \cite{Wojcik2009} have derived an extended model based on the Onsager theory which leads to a weaker field dependence.

\subsection{Transient current sensitivity}

The sensitivity analysis in thickness dependence and current-voltage curves have shown, that several input parameters have a similar influence on the output and thus those parameters can not be determined from these measurements. In this part transient simulations are discussed and their sensitivity on parameters is analyzed.\\
\begin{figure}
\begin{center}
\resizebox{8.5cm}{!}{\includegraphics{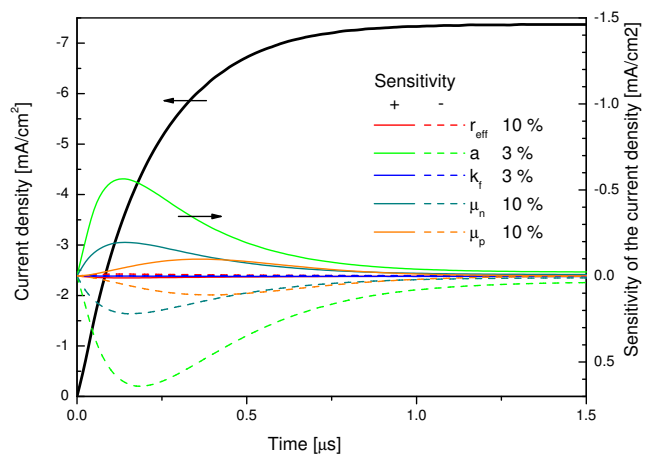}}
\end{center}
\caption{ \label{fig:turn_on} Sensitivity  of the short-circuit current during turn-on.}
\end{figure}
Figure \ref{fig:turn_on} shows the simulated transient of the short-circuit current when the light is turned on at t=0, for a thickness of the active layer of 70~nm. Again the parameter set shown in table \ref{tab:parameters1} and \ref{tab:parameters2} has been used. The sensitivity clearly shows that in the turn-on dynamics the Langevin recombination efficiency $r_{eff}$ and the decay rate $k_f$ do not play an important role, therefore the turn-on dynamics is solely defined by the pair separation distance $a$ and the two mobilities. The CT-exciton decay rate and thus the recombination of free charge carriers play no role. The electron mobility determines the first part of the turn-on and the sensitivity reaches its maximum at 0.15~$\mu s$,  whereas the hole mobility has its maximum at 0.35~$\mu s$. Further analysis (not shown) reveals that the electron mobility $\mu_n$ has exactly the same influence on the turn-on behavior as does the pair separation a. This indicates that the turn-on behavior is limited by the time it takes to dissociate the CT-exciton pair.  When the CT-exciton parameters are chosen such that the average lifetime of the CT-exciton is much smaller than the time it takes to extract the charge carriers, the turn-on behavior is solely defined by the two mobilities.\\
\begin{figure}
\begin{center}
\resizebox{8.5cm}{!}{\includegraphics{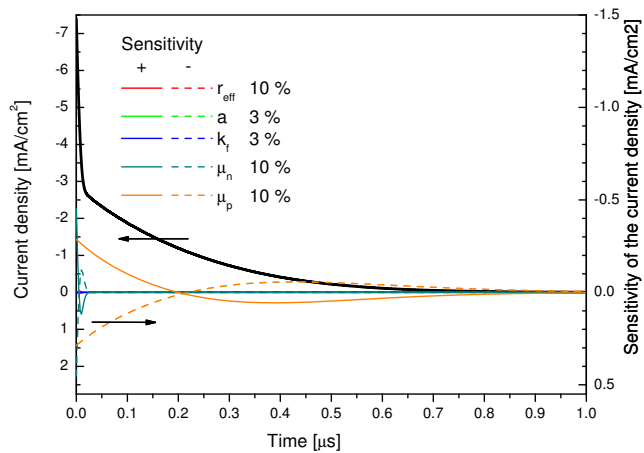}}
\end{center}
\caption{ \label{fig:turn_off} Sensitivity  of the short-circuit current during turn-off.}
\end{figure} 
Figure \ref{fig:turn_off} shows the transient of the short-circuit current when the light is turned off at t=0. The shape of the transient turn-off curve shown in figure \ref{fig:turn_off} is clearly different from the turn-on curve shown in figure  \ref{fig:turn_on}. The first one has a clearly visible kink at t = 0.015 $\mu s$ whereas the latter curve is more smooth without any visible kink.  The sensitivity analysis of the turn-off behavior shows that the curve is solely dependent on the two mobilities. This means that CT-exciton and recombination parameters play no role and thus the recombination and subsequent CT-exciton dissociation play no role. The initial drop is given by the mobility of the electrons $\mu_n$ whose influence peaks at t=0.01 $\mu s$ whereas the second, much slower drop is given by the hole mobility $\mu_p$ whose influence peaks at t=0.45 $\mu s$. This difference is in the same range as the difference in the mobilities of the two charge carriers which is $\frac{\mu_n}{\mu_p}=50$.\\
This last sensitivity analysis shows, that by measuring the short-circuit current during turn-off, the two mobility parameters can be estimated. This observation is similar to the findings by Pinner et al. \cite{Pinner:1999p1243} for electroluminescence transients in organic light-emitting diodes where the turn-off regime is governed by the built-in field as well. The time it takes to extract all charge carriers and thus the mobilities can be calculated under the assumption that the active layer is homogeneously filled with charge carriers and that the internal electric field is given by the difference on the applied and the built-in voltage divided by the active layer thickness $E=\frac{(V-V_{bi})}{L}$. We thus obtain

\begin{equation}
\mu = \frac{L^2}{(V_{appl}-V_{bi})\tau}.
\end{equation}

The parameter  $\tau$ is the time it takes for the current to drop down to 5\% of the initial value. With this approach the electron as well as the hole mobility can be extracted separately. 

\begin{figure}
\begin{center}
\resizebox{8cm}{!}{\includegraphics{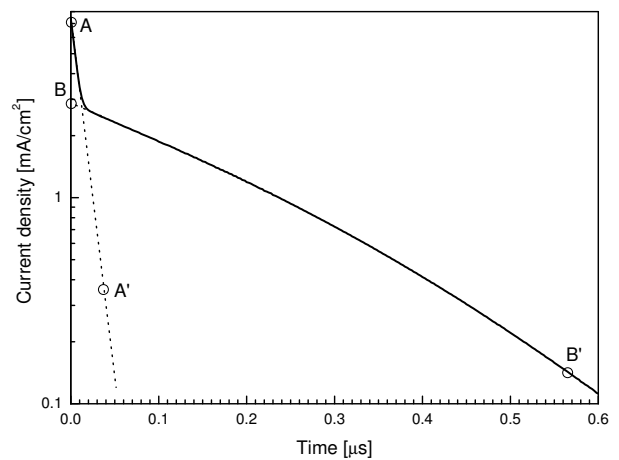}}
\end{center}
\caption{ \label{fig:turn_off_log} Semi-logarithmic plot of the turn-off behavior shown in figure  \ref{fig:turn_off}. This plot can be used to extract the mobilities of electrons and holes.}
\end{figure} 

Figure \ref{fig:turn_off_log} shows the turn-off behavior as a semi-logarithmic plot. The first steep drop of the current is associated with the electrons and thus a line is fitted to it. Point A' is defined as the point along the fitted curve were the initial current at point A is reduced to 5\%. The same is done for the second drop which is associated with the holes. Again B' is defined as the point where the current dropped down to 5\% with B as the reference point. This gives for  $\tau_n$=0.037~$\mu$s and $\tau_p$=0.56~$\mu$s. With V$_{appl}$-V$_{bi}$=0.69~V (cf. table \ref{tab:parameters1}) and an active layer thickness 70~nm the following mobilities are extracted:  $\mu_{n}$ = 1.9$\times$10$^{-7}$ m$^2$/Vs and $\mu_{p}$ = 1.3$\times$10$^{-8}$ m$^2$/Vs. The extracted values are in the same order of magnitude as the input values given in table \ref{tab:parameters1}. The mobility of the fast charge carriers is underestimated and the mobility of the slow charge carriers is slightly overestimated. This discrepancies stem from the assumption of a homogeneous electric field inside the device and thus the independent treatment of electrons and holes which is a simplification.

\section{Conclusion}

From the measured thickness-dependent short-circuit current for thin active layers and the optical model we estimated a photon to CT-exciton conversion efficiency $g_{eff}$ of 66\%. Using this dataset the coupled opto-electronic simulation suggests a lower limit for the CT-exciton dissociation efficiency of 90\%. Adding the measured current-voltage curve to the numerical analysis and assuming that $k_{f}$ is equal to 1$\times$10$^{5}$ s$^{-1}$ we find dissociation parameters that are consistent with the estimated 90\% dissociation efficiency. From a comparison of several current-voltage curves at different thicknesses there is evidence that the influence of the electric field on the CT-exciton dissociation process is overestimated using the Onsager-Braun model. Further investigations have to be carried out to clarify this issue. The sensitivity analysis conducted demonstrates that the influence of the two exciton parameters and the electron mobility are linearly dependent in the current-voltage curve and photocurrent thickness scaling. A sensitivity analysis of transient processes indicates that the input parameters can be separated. During turn-on, the CT-exciton decay rate $k_f$ and thus the charge carrier recombination play no role. Even more insight can be gained by looking at the turn-off behavior. In this regime the transient current density is solely defined by the mobilities of the charge carriers. This suggests that the charge carrier mobilities can be extracted. The influence of charge traps was not considered in this study but will be discussed elsewhere.

\section{Acknowledgment}

We would like to thank Jan Gilot and Prof. René A. J. Janssen from TU Eindhoven for supplying us with a consistent set of measurements and helpful discussions.
The authors gratefully acknowledge financial support of the Swiss Federal Office of Energy.

\bibliography{Sensitivity}

\end{document}